\documentclass[longauth]{aa} % for the long lists of affiliations
\usepackage{graphicx,natbib}
\begin{document}
   \title{The unprecedented optical outburst of the quasar \object{3C 454.3}}

   \subtitle{The WEBT campaign of 2004--2005\thanks{For questions
   regarding the availability of the data presented in this paper,
   please contact Massimo Villata ({\tt villata@to.astro.it}).}}

\author{M.~Villata \inst{1}
\and C.~M.~Raiteri \inst{1}
\and T.~J.~Balonek \inst{2}
\and M.~F.~Aller \inst{3}
\and S.~G.~Jorstad \inst{4}
\and O.~M.~Kurtanidze \inst{5,6,7}
\and F.~Nicastro \inst{8}
\and K.~Nilsson \inst{9}
\and H.~D.~Aller \inst{3}
\and A.~Arai \inst{10}
\and A.~Arkharov \inst{11}
\and U.~Bach \inst{1}
\and E.~Ben\'{\i}tez \inst{12}
\and A.~Berdyugin \inst{9}
\and C.~S.~Buemi \inst{13}
\and M.~B\"ottcher \inst{14}
\and D.~Carosati \inst{15}
\and R.~Casas \inst{16}
\and A.~Caulet \inst{12}
\and W.~P.~Chen \inst{17}
\and P.-S.~Chiang \inst{17}
\and Y.~Chou \inst{17}
\and S.~Ciprini \inst{9,18}
\and J.~M.~Coloma \inst{16}
\and G.~Di~Rico \inst{19}
\and C.~D\'{\i}az \inst{20}
\and N.~V.~Efimova \inst{11,21}
\and C.~Forsyth \inst{2}
\and A.~Frasca \inst{13}
\and L.~Fuhrmann \inst{1,18}
\and B.~Gadway \inst{2}
\and S.~Gupta \inst{14}
\and V.~A.~Hagen-Thorn \inst{21,22}
\and J.~Harvey \inst{14}
\and J.~Heidt \inst{7}
\and H.~Hernandez-Toledo \inst{12}
\and F.~Hroch \inst{23}
\and C.-P.~Hu \inst{17}
\and R.~Hudec \inst {24}
\and M.~A.~Ibrahimov \inst{25}
\and A.~Imada \inst{26}
\and M.~Kamata \inst{10}
\and T.~Kato \inst{26}
\and M.~Katsuura \inst{10}
\and T.~Konstantinova \inst{21}
\and E.~Kopatskaya \inst{21}
\and D.~Kotaka \inst{10}
\and Y.~Y.~Kovalev \inst{27,28}
\and Yu.~A.~Kovalev \inst{28}
\and T.~P.~Krichbaum \inst{29}
\and K.~Kubota \inst{26}
\and M.~Kurosaki \inst{10}
\and L.~Lanteri \inst{1}
\and V.~M.~Larionov \inst{21,22}
\and L.~Larionova \inst{21}
\and E.~Laurikainen \inst{30}
\and C.-U.~Lee \inst{31}
\and P.~Leto \inst{32}
\and A.~L\"ahteenm\"aki \inst{33}
\and O.~L{\'o}pez-Cruz \inst{34}
\and E.~Marilli \inst{13}
\and A.~P.~Marscher \inst{4}
\and I.~M.~McHardy \inst{35}
\and S.~Mondal \inst{17}
\and B.~Mullan \inst{2}
\and N.~Napoleone \inst{36}
\and M.~G.~Nikolashvili \inst{5}
\and J.~M.~Ohlert \inst{37}
\and S.~Postnikov \inst{14}
\and T.~Pursimo \inst{38}
\and M.~Ragni \inst{19}
\and J.~A.~Ros \inst{16}
\and K.~Sadakane \inst{10}
\and A.~C.~Sadun \inst{39}
\and T.~Savolainen \inst{9}
\and E.~A.~Sergeeva \inst{40}
\and L.~A.~Sigua \inst{5}
\and A.~Sillanp\"a\"a \inst{9}
\and L.~Sixtova \inst{23}
\and N.~Sumitomo \inst{10}
\and L.~O.~Takalo \inst{9}
\and H.~Ter\"asranta \inst{33}
\and M.~Tornikoski \inst{33}
\and C.~Trigilio \inst{13}
\and G.~Umana \inst{13}
\and A.~Volvach \inst{41}
\and B.~Voss \inst{42}
\and S.~Wortel \inst{2}
}

   \offprints{M.\ Villata}

\institute{INAF, Osservatorio Astronomico di Torino, Italy 
\and Foggy Bottom Observatory, Colgate University, NY, USA
\and Department of Astronomy, University of Michigan, MI, USA
\and Institute for Astrophysical Research, Boston University, MA, USA
\and Abastumani Astrophysical Observatory, Georgia
\and Astrophysikalisches Institut Potsdam, Germany
\and Landessternwarte Heidelberg-K\"onigstuhl, Germany
\and Harvard-Smithsonian Center for Astrophysics, MA, USA 
\and Tuorla Observatory, Finland
\and Astronomical Institute, Osaka Kyoiku University, Japan
\and Main (Pulkovo) Astronomical Observatory of the Russian Academy of Sciences, Russia
\and Instituto de Astronom\'{\i}a, UNAM, Mexico 
\and INAF, Osservatorio Astrofisico di Catania, Italy
\and Astrophysical Institute, Department of Physics and Astronomy, Ohio University, OH, USA
\and Armenzano Astronomical Observatory, Italy
\and Agrupaci\'o Astron\`omica de Sabadell, Spain
\and Institute of Astronomy, National Central University, Taiwan
\and Dipartimento di Fisica e Osservatorio Astronomico, Universit\`a di Perugia, Italy
\and INAF, Osservatorio Astronomico di Teramo, Italy
\and Departamento de Astrof\'{\i}sica, Universidad Complutense, Spain
\and Astronomical Institute, St.-Petersburg State University, Russia
\and Isaac Newton Institute of Chile, St.-Petersburg Branch
\and Institute of Theoretical Physics and Astrophysics, Masaryk University, Czech Republic
\and Astronomical Institute, Academy of Sciences of the Czech Republic, Czech Republic
\and Ulugh Beg Astronomical Institute, Academy of Sciences of Uzbekistan, Uzbekistan
\and Department of Astronomy, Kyoto University, Japan
\and Jansky fellow, National Radio Astronomy Observatory, WV, USA
\and Astro Space Center of Lebedev Physical Institute, Russia
\and Max-Planck-Institut f\"ur Radioastronomie, Germany
\and Division of Astronomy, University of Oulu, Finland
\and Korea Astronomy and Space Science Institute, South Korea
\and INAF, Istituto di Radioastronomia Sezione di Noto, Italy
\and Mets\"ahovi Radio Observatory, Helsinki University of Technology, Finland
\and Instituto Nacional de Astrof\'{\i}sica, Optica y Electr\'onica (INAOE), Mexico
\and School of Physics and Astronomy, The University, UK
\and INAF, Osservatorio Astronomico di Roma, Italy
\and Michael Adrian Observatory, Germany
\and Nordic Optical Telescope, Roque de los Muchachos Astronomical Observatory, TF, Spain
\and Department of Physics, University of Colorado at Denver and Health Sciences Center, CO, USA
\and Crimean Astrophysical Observatory, Ukraine
\and Radio Astronomy Laboratory of Crimean Astrophysical Observatory, Ukraine
\and Institut f\"ur Theoretische Physik und Astrophysik der Universit\"at Kiel, Germany
}

   \date{}

% \abstract{}{}{}{}{} 
% 5 {} token are mandatory
 
  \abstract
   {The radio quasar \object{3C 454.3} underwent an exceptional optical outburst lasting more than 1 year and
   culminating in spring 2005. The maximum brightness detected was $R=12.0$, which represents
   the most luminous quasar state thus far observed ($M_B \sim -31.4$).}
   {In order to follow the emission behaviour of the source in detail, a large multiwavelength
   campaign was organized by the Whole Earth Blazar Telescope (WEBT).}
   {Continuous optical, near-IR and radio monitoring was performed in several bands. 
   ToO pointings by the Chandra and INTEGRAL satellites provided additional information at high
   energies in May 2005.}
   {The historical radio and
   optical light curves show different behaviours. Until about 2001.0 only moderate
   variability was present in the optical regime, while prominent and long-lasting radio
   outbursts were visible at the various radio frequencies, with higher-frequency variations
   preceding the lower-frequency ones. After that date, the optical activity increased
   and the radio flux is less variable. This suggests that the optical and radio emissions
   come from two separate and misaligned jet regions, with the inner optical one acquiring 
   a smaller viewing angle during the 2004--2005 outburst. 
   Moreover, the colour-index behaviour (generally redder-when-brighter) during the outburst
   suggests the presence of a luminous accretion disc. 
   A huge mm outburst followed the optical one, peaking in June--July 2005.
   The high-frequency (37--43 GHz) radio flux started to increase in early 2005 and reached a maximum at
   the end of our observing period (end of September 2005). VLBA observations at 43 GHz during the summer 
   confirm the brightening of the radio core and show an increasing polarization.
   An exceptionally bright X-ray state was detected in May 2005, corresponding to the rising mm flux and 
   suggesting an inverse-Compton nature of the hard X-ray spectrum.}
   {A further multifrequency monitoring effort is needed to follow the next phases of this 
   unprecedented event.}


   \maketitle
%
%________________________________________________________________

\section{Introduction}
The flat-spectrum and highly polarized radio quasar \object{3C 454.3} 
at $z=0.859$ belongs to the blazar class
of active galactic nuclei. The dominant non-thermal emission of blazars is ascribed
to a plasma jet oriented at a small angle to the line of sight, so that the radiation
is relativistically beamed.
The broad-band spectral energy distribution (SED) of blazars shows two components: 
the lower-energy one is attributed to synchrotron radiation by relativistic electrons, while the
higher-energy one is commonly thought to be due to inverse-Compton scattering
of soft photons by the same relativistic electrons. 
   \begin{figure*}
   \centering
   \includegraphics{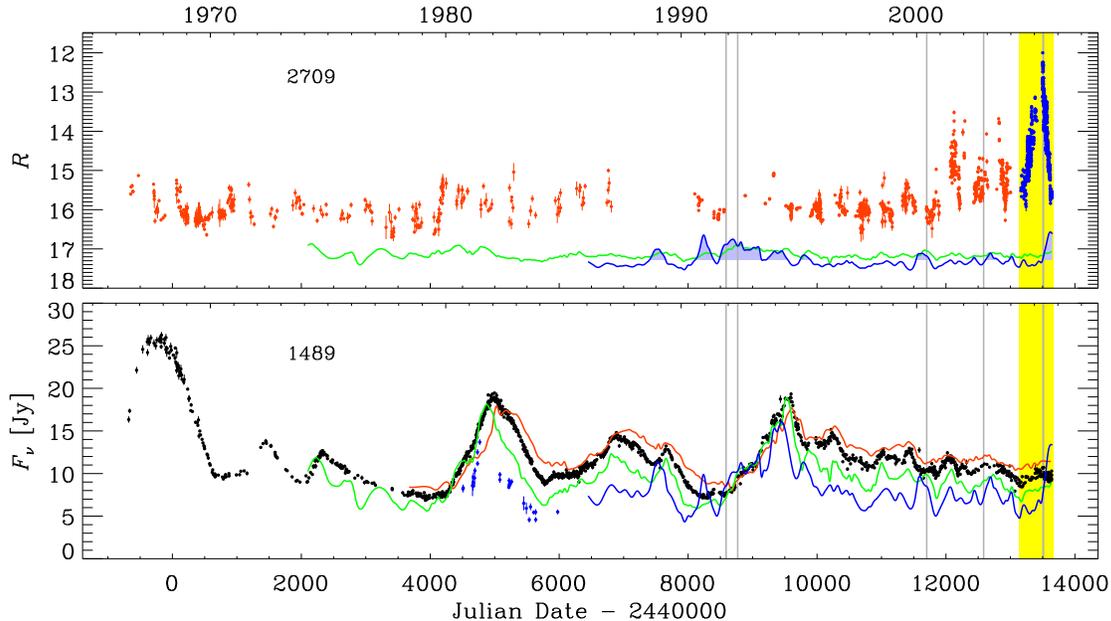}
      \caption{Optical (top) and radio (bottom) light curves from 1966 to the end of September 2005;
      the yellow strip indicates the period of the 2004--2005 WEBT campaign, while grey
      vertical lines show the times of the satellite pointings (see Fig.\ \ref{sed}).
      Bottom panel: black points indicate the 8 GHz light curve (1489 data points);
      red, green, and blue curves represent the cubic spline interpolations through the
      30-day binned light curves at 5, 14.5, and 37 GHz, respectively;
      blue points are sparse 37 GHz data.
      Top panel: besides the $R$-band light curve (2709 data points, 1502 of which are from 
	the WEBT campaign, blue points), the radio hardness ratios
      $H_{37/8}$ (blue line) and $H_{14.5/8}$ (green line) are plotted as $18-H$; the light blue
      shading highlights where $H_{37/8}$ is harder than average.}
         \label{radop}
   \end{figure*}

In addition to these two components,
quasars can also show a ``big blue bump" in the UV band, which is interpreted
as the signature of the thermal disc feeding the supermassive black hole.
This component is mostly visible when the synchrotron emission is faint \citep[see e.g.][]{hag94,pia99}.

One of the main features of blazars is their strong emission variability at all wavelengths,
from the radio band to $\gamma$-ray energies, on time scales ranging
from a few minutes to several years.

\section{The WEBT campaign}
The historical optical light curve of 3C 454.3 starts from $\sim 1900$.
In the radio bands the source has been monitored since the 1960s.
Figure \ref{radop} shows the optical (top) and radio (bottom) light curves over the last $\sim 40$ years.
The historical radio data are from the University of Michigan Radio Astronomy Observatory (UMRAO) and
the Mets\"ahovi Radio Observatory (\citealt{ter05}, references therein, and unpublished data), 
while data from the Crimean (RT-22), MDSCC (PARTNeR), Medicina, Noto, and
SAO RAS (RATAN-600) Observatories as well as data from the VLA/VLBA Polarization Calibration 
Database\footnote{{\tt http://www.vla.nrao.edu/astro/calib/polar/}.}
contribute to the most recent part of the light curves.
In the figure the radio data are represented as black points (8 GHz), red line (cubic spline through the
30-day binned 5 GHz data), green line (the same for 14.5 GHz), while blue points/spline indicate the 37
GHz fluxes (including also 36 GHz data).
The radio light curves show prominent, long-lasting outbursts, with
higher-frequency variations leading the lower-frequency ones, as typically found in many blazars.

The optical light curve shows two very different phases.
Until $\sim 2001.0$ only moderate variability is observed in the range $R \sim 15$--17 
(an exception being the higher activity shown in $\sim 1940$--1955, when $R$ reached $\sim 14$; 
see \citealt{ang68}).
From 2001 the amplitude of the variations starts to increase.

\begin{figure*}
   \centering
   \includegraphics{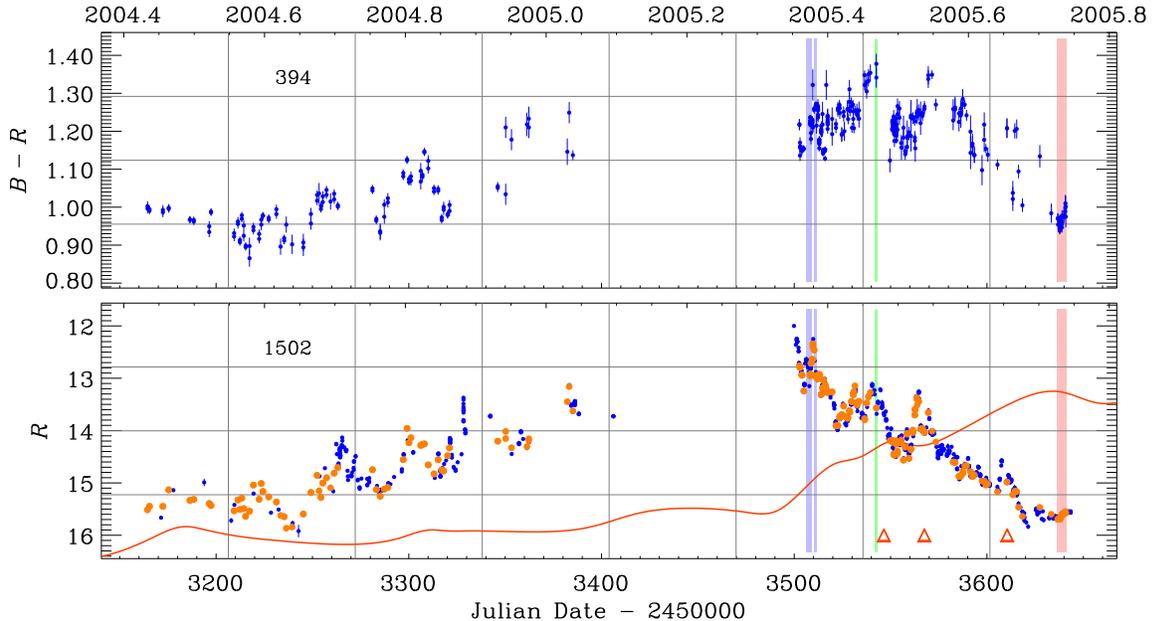}
      \caption{Time evolution of the $B-R$ colour index (top) during the WEBT campaign
      compared with the $R$-band light curve (bottom); orange points in the
      light curve highlight the $R$ data used to derive the $B-R$ indices;
      the red curve is a cubic spline interpolation through the 43 GHz light curve, 
      arbitrarily scaled to fit the figure; the light blue strips indicate the epochs 
      of the INTEGRAL (left) and Chandra (right) pointings, while the light green and red ones 
      refer to spectra shown in Fig.\ \ref{sed}; the red triangles mark the VLBA epochs 
      of Fig.\ \ref{maps}.}
         \label{colori}
   \end{figure*}

On May 9.36, 2005 (at the start of the observing season) the source was observed at $R=12.0$\footnote{
This outburst peak likely represents the most luminous quasar state ever observed: $M_{B} \sim -31.4$
($H_0=70$ km $\rm s^{-1}$ $\rm Mpc^{-1}$, $\Omega_M=0.3$, $\Omega_\Lambda=0.7$), comparable only
to the 3C 279 outburst of 1937 ($M_{B} \sim -31.2$). 
For comparison, the bright but nearby quasar 3C 273
has ``only" $M_{B} \sim -26.4$, while the most luminous quasars of the SDSS and of the $z>4$ sample 
of \citet{vig03} reach $M_{B} \sim -29.1$ and $M_{B} \sim -30.2$, respectively.},
thus triggering a  multifrequency campaign by the Whole Earth Blazar Telescope
(WEBT)\footnote{{\tt http://www.to.astro.it/blazars/webt/}; see e.g.\ \citet{vil04a,rai05}.}.
In order to have information on both the rising and decreasing phases of the outburst,
data were collected from June 2004 to the end of September 2005.
In total, 5584 $UBVRI$ observations from 18 telescopes\footnote{Kyoto, 
Osaka Kyoiku, Lulin, Mt.\ Maidanak, Abastumani, Crimean, MonteBoo,
Catania, Armenzano, Michael Adrian, Torino, Sabadell, Teide (BRT),
Roque de los Muchachos (KVA and Liverpool), Foggy Bottom, Kitt Peak (MDM), and 
San Pedro Martir Observatories.}
were performed in this period;
moreover, $JHK$ data were taken at Campo Imperatore, Calar Alto, and Roque de los Muchachos (NOT).
Radio data from 1 to 43 GHz were acquired at several telescopes, mentioned above.
In the top panel of Fig.\ \ref{radop} the $R$-band light curve from the WEBT campaign
is displayed as blue dots (see also Fig.\ \ref{colori}): it is composed of 1502 points. In particular,
1146 data points were taken in 143.59 days,
from May 9.36 to September 29.95, 2005, during the outburst decreasing phase,
with a mean time separation of 3.0 hours, and
only 4 gaps longer than 36 hours (2--4 days). 
Previous data are shown as red points; they come from the literature (including the recent
paper by \citealt{fuh06}) as well as from some WEBT observatories.
Here we present some results of this campaign;
a more detailed study is deferred to a forthcoming paper.

By looking at Fig.\ \ref{radop} no evident correlation appears between the 
optical and radio events. In particular, a low radio state characterizes 
the 2004--2005 optical outburst as well as
the previous optical activity, while the past strong radio 
outbursts do not show any optical counterpart. Even the hardness ratio 
between the 37 GHz and 8 GHz splines ($H_{37/8}=F_{37}/F_8$, plotted in the top panel as $18-H_{37/8}$, 
blue line) does not show correlation with the optical, as is found by \citet{vil04b} for BL Lac.
Only during the WEBT campaign can one see a relation:
a fast rise of $H_{37/8}$ to its maximum value (1.43),
delayed by several months with respect to the rise of the optical outburst.
This growth of $H_{37/8}$ is essentially due to the increase of the 37 GHz flux
visible in the bottom panel. The 43 GHz light curve, 
not plotted there (but in Fig.\ \ref{colori}, see below), confirms this trend.
In the top panel $H_{14.5/8}$ is also plotted (green line) for comparison.
In general, its trend mimics fairly well that of $H_{37/8}$, with some delay and
reduced variability amplitude. In particular, just a slight increase can be seen in 2005.
The 22 GHz flux and the corresponding $H_{22/8}$ display an intermediate trend, as expected.
   \begin{figure}
   \centering
   \includegraphics[height=13cm]{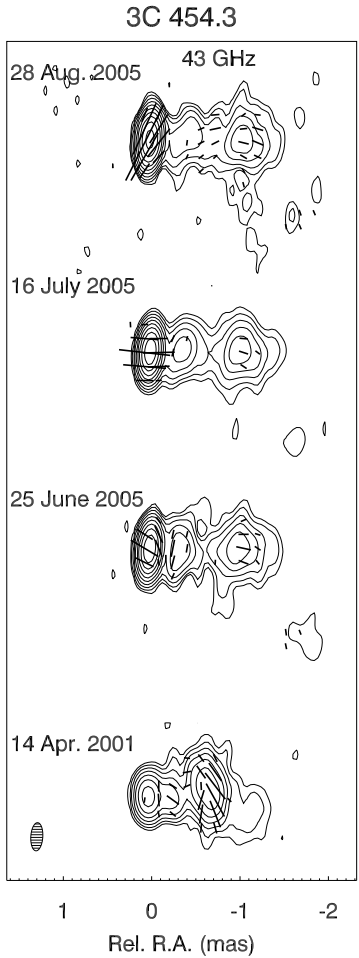}
      \caption{VLBA maps at 43 GHz; the total intensity is shown by contours from
0.15\% up to 76.8\% of the August 2005 peak value, increasing by a factor of 2;
the straight lines show the electric vector and their length is proportional to the local 
polarized intensity.}
         \label{maps}
   \end{figure}

The general lack of correlation between the radio and optical behaviour suggests that
the optically emitting region is not transparent to the radio frequencies, so that these must come from 
another, outer part of the jet, possibly misaligned with respect to the former. It seems that before 2001 
the radio part was more aligned with the line of sight and the radio variability was thus enhanced 
by Doppler effects. On the contrary, 
in the last 5 years the optical radiation would dominate the scenario due to a 
smaller viewing angle of the corresponding emitting region.
Thus the flux increase we are seeing at the high radio frequencies should be produced 
close to the optical region. The outburst will probably
propagate further out, towards lower-frequency emitting regions.
However, we cannot foresee whether it will be
visible, since this depends on the jet curvature.

A confirmation of the above general picture comes from the source behaviour in the mm bands, where
a strong outburst was observed to peak in June--July 2005, reaching $\sim 45$ Jy at 1 mm and $\sim 
26$ Jy at 3 mm 
(data from the IRAM 30-m telescope), while previous observations in 1985--2004 did not show any comparable 
activity, the flux density ranging between $\sim 2$ and $\sim 15$ Jy in both bands 
\citep[data from][ and unpublished data from IRAM and SEST]{ter92,ter98,ter04,ste93,tor96,reu97}.

The time evolution of the $B-R$ colour index during the optical outburst is shown in Fig.\
\ref{colori} (top panel), compared with the $R$-band light curve (bottom panel).
Orange dots in the latter panel highlight the $R$-band points used
to derive the $B-R$ colour indices.
In general, a ``redder when brighter" trend can be recognized.
Since 3C 454.3 is a quasar, this can be interpreted
as due to the contribution given by the thermal emission from the accretion disc, 
which mainly affects the bluer region of the optical spectrum when the jet emission is faint.
In other words, there would be two optical components: a variable one from the jet and the other
from the disc, not necessarily variable. When the jet is brighter, its redder spectrum dominates, 
and vice versa.
However, there seems to be a kind of ``saturation" effect in the $B-R$ trend:
in the brightest part of the outburst ($R \la 14$) the trend is not visible any longer.
Most likely, this is due to the dominance of the non-thermal jet emission,
which would display its usual ``bluer when brighter" behaviour, 
thus balancing the opposite trend.

In the bottom panel of Fig.\ \ref{colori} the 43 GHz flux-density spline is also shown 
(arbitrarily scaled to fit the figure, the range is $\sim 4$--13 Jy) for comparison with
the optical light curve. It peaks in September 2005, i.e.\ 2--3 months 
after the mm outburst, which seems to have a similar delay with respect to the optical event.

\section{VLBA observations}
Figure \ref{maps} shows three recent VLBA maps at 43 GHz compared with an older map
dated April 2001\footnote{Multiepoch and multifrequency VLBA monitoring of 3C 454.3 
(with polarization) was performed between May and December 2005 (Savolainen et al., in preparation).}.
The plot is constructed with both total and polarized intensity, normalized
to the peak values achieved at the epoch of August 2005 ($S_{\rm tot}=9.62$ Jy/beam,
$S_{\rm pol}=0.162$ Jy/beam).
The total intensity is shown by contours from
0.15\% up to 76.8\% of the August 2005 peak value, increasing by a factor of 2.
The straight lines show the
electric vector and their length is proportional to the local polarized intensity.

There is a significant increase in both total and polarized intensity of the VLBI core
compared with the values in April 2001 (see also Table \ref{vlba}), thus confirming the expectation
that the 43 GHz rising flux in 2005 seen in Fig.\ \ref{colori} comes from the VLBI core.
The three VLBA epochs are indicated in Fig.\ \ref{colori} as red triangles: there is a strict agreement 
between the trend of the 43 GHz light curve and the total intensity values reported in Table \ref{vlba}.
Since the VLBI core of 3C 454.3 at 43 GHz used to have very low polarization
compared to the western feature at $\sim 0.7$ mas from the core, a strong increase
of the core polarization along with the total flux suggests
a new component emerging from the core. However, as discussed in the previous section,
the emerging feature will be more or less visible depending on the jet bending.
\begin{table}
\caption{Results of VLBA observations of 3C 454.3 at 43 GHz.}
\label{vlba}
\centering
\begin{tabular}{l c c }
\hline\hline
Epoch    &  Total Int.\ (mJy)  &  Polarized Int.\ (mJy)\\
\hline
April 14, 2001  &     1710    &       15  \\
June 25, 2005   &     6560    &       74  \\
July 16, 2005   &     6290    &      103  \\
August 28, 2005 &    10200    &      200  \\
\hline
\end{tabular}
\end{table}
   \begin{figure*}
%   \sidecaption
   \centering
   \includegraphics[width=13cm]{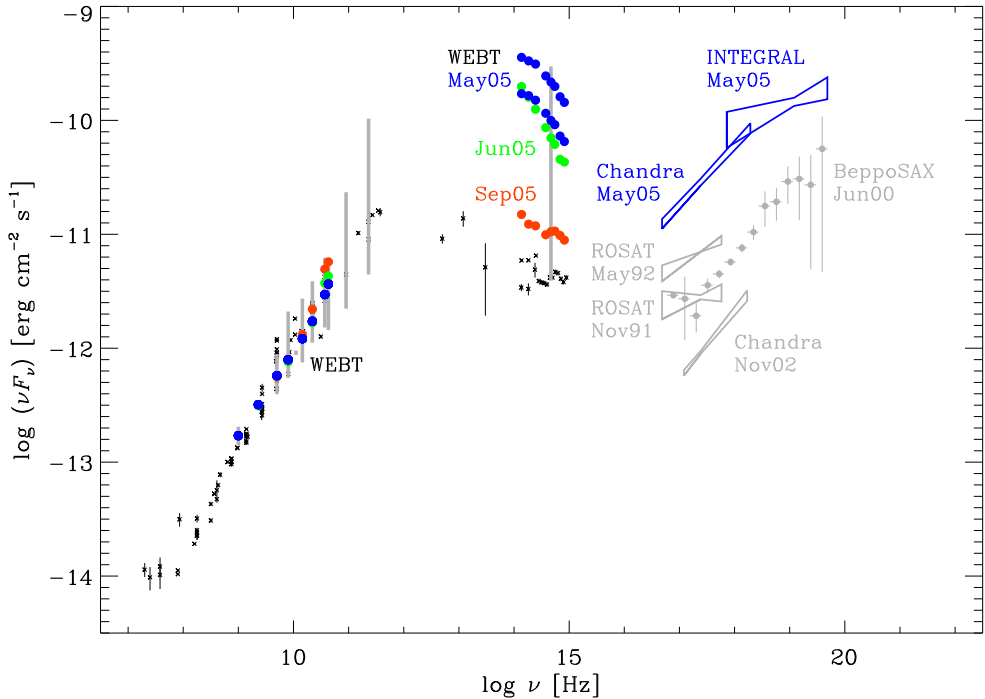}
      \caption{Spectral energy distribution of 3C 454.3 showing contemporaneous radio, near-IR, optical,
      and X-ray (Chandra and INTEGRAL) data during May 15--20, 2005. Previous data are also
      plotted for comparison, together with two other spectra from the WEBT campaign (see text for 
      further details).}
         \label{sed}
   \end{figure*}

\section{Chandra observation}
3C 454.3 was observed by Chandra on May 19.7--21.0, 2005 with the HRC-LETG
instrumental configuration. The Chandra pointing was performed as
part of a ToO program on blazars in outburst, and lasted 112 ks.
We used version 3.3 of the CIAO software to reduce and analyse the data.
We extracted source and background spectra from the standard
HRC-LETG {\em bow-tie} regions, and then grouped the source spectrum to contain
at least 20 counts per channel. We used the fitting package {\tt Sherpa}
to fit the background-subtracted source spectrum in the range $\sim 0.2$--8 keV.
Here we report the results of the continuum spectral analysis, while 
the study of narrow absorption features due to possible intervening Warm-Hot
Intergalactic Medium filaments is deferred to a forthcoming paper (Nicastro et al.,
in preparation).

The 0.2--8 keV spectrum is well described ($\chi^2/{\rm d.o.f.} = 1817/2368$)
by a rather flat power law, with photon index $\Gamma = 1.477 \pm 0.017$, absorbed by a large
column density $N_{\rm H} = (13.40 \pm 0.05) \times 10^{21}$ $\rm cm^{-2}$,
exceeding by more than a factor of 2 the Galactic value.
We measured de-absorbed fluxes
of $F_{0.2-2} = (5.5 \pm 0.2) \times 10^{-11}$ erg $\rm cm^{-2}$ $\rm s^{-1}$ and
$F_{2-8} = (8.4 \pm 0.2) \times 10^{-11}$ erg $\rm cm^{-2}$ $\rm s^{-1}$.

\section{Spectral energy distribution and conclusions}
In Fig.\ \ref{sed} we show the broad-band SED of the source.
The small black crosses in the radio--optical range indicate archival data taken from NED.
In the X-ray band we plotted data from old
observations by ROSAT in November 1991 \citep{sam97} and in May 1992 \citep{pri96},
as well as data from a BeppoSAX observation in June 2000\footnote{{\tt http://www.asdc.asi.it/bepposax/}.},
and data from a pile-up affected Chandra observation in November 2002
\citep{mar05}.

During the brightest phases of the optical outburst, in May 2005,
there were other observations by high-energy satellites (RXTE, Swift, INTEGRAL)
besides the Chandra one presented in the previous section.
The ToO INTEGRAL observation was performed on May 15.8--18.4.
In Fig.\ \ref{sed} we show the corresponding 3--200 keV spectrum \citep{pia06}.
The Chandra spectrum of May 19.7--21.0 is also plotted, together with radio, near-IR and
optical data from the WEBT in the period of the two pointings (blue symbols).
Since in this period the optical brightness varied by $\Delta R = 0.9$,
we plot two $UBVRIJHK$ spectra (each composed of simultaneous data) corresponding to a high and a low state. 
No significant variation 
in the spectral slope is seen between the two states, despite their different brightness levels, in
agreement with the rather stable colour index found in this period (see Fig.\ \ref{colori}, light blue strips).
The contemporaneous radio data follow a power law from 1 to 22 GHz, which then breaks due to the
37 and 43 GHz fluxes that begin to rise (see Figs.\ \ref{radop} and \ref{colori}). The grey
vertical strips in Fig.\ \ref{sed} show the historical variation range of the radio, mm and 
$R$-band data collected for this paper (and partly displayed in Figs.\ \ref{radop} and \ref{colori}).

One month after the May satellite pointings, around June 21.0, the reddest observed optical state is
achieved (light green line in Fig.\ \ref{colori}). The corresponding near-IR--optical spectrum is
displayed in Fig.\ \ref{sed} with green symbols: it is indeed very steep, crossing the lower May spectrum
around the $H$ band, and most likely connecting with the highest parts of the mm strips, since
we should be close to the mm outburst peak. The 37--43 GHz flux has also increased 
(see Figs.\ \ref{radop} and \ref{colori}), while no appreciable variation is seen at lower frequencies 
(where different-colour symbols often overlap in the figure, so that only the blue circles are visible).

In late September, at the foot of the outburst, the bluest optical state of 2005 is found 
(light red strip in Fig.\ \ref{colori}). The corresponding (red) spectrum in Fig.\ \ref{sed} is the average 
over 5 days (September 23.0--28.0) of intensive $UBVRIJHK$ monitoring: the thermal component seems to strongly
affect the spectrum. At this time the 22 GHz flux starts to increase and the radio spectral break shifts 
downwards,
while a peak of the outburst is seen at 37--43 GHz (see also Figs.\ \ref{radop} and \ref{colori}).

The main feature of the plotted SEDs is the strong variability of the optical--IR, mm
and X-ray fluxes.
In Fig.\ \ref{radop} vertical grey lines indicate the times of the X-ray pointings.
The X-ray flux seems to correlate neither with radio flux (always rather low), 
nor with the optical level
(in 2002 it was brighter than during the previous pointings, but the X-ray flux was lower),
nor with the radio hardness (higher during the ROSAT pointings and intermediate during the others).
It seems instead to correlate with the mm flux, since a strong mm outburst was close to its peak
during the May 2005 X-ray observations. This is expected
due to the probable inverse-Compton nature of the hard X-ray spectrum.

Thus, this unprecedented quasar optical outburst preceded an equally unprecedented mm outburst (accompanied
by an exceptionally bright X-ray state) by 2--3 months, and, 2--3 months later, it had fully propagated to the
high radio frequencies. This occurred in the VLBI radio core. We cannot foresee whether and when
further propagation will be visible at lower frequencies and outside the radio core,
due to probable jet bending. Further multifrequency monitoring 
is needed to follow the next phases of this exceptional event.

\begin{acknowledgements}
The mm data from the IRAM 30-m telescope were provided by H.\ Ungerechts
based on preliminary results from regular flux monitoring observations
by the IRAM Granada staff. 
This work is partly based on observations made with the Nordic Optical Telescope, operated
on the island of La Palma jointly by Denmark, Finland, Iceland,
Norway, and Sweden, in the Spanish Observatorio del Roque de los
Muchachos of the Instituto de Astrof\'{\i}sica de Canarias. It is partly based also
on observations collected at the German-Spanish Astronomical 
Center (DSAZ), Calar Alto, operated by the Max-Planck-Institut 
f\"ur Astronomie Heidelberg jointly with the Spanish National 
Commission for Astronomy.
We thank Calar Alto for allocation director's discretionary time 
to this programme.
This research has made use of data from the University of Michigan Radio Astronomy Observatory,
which is supported by the National Science Foundation and by funds from the University of Michigan,
and of the NASA/IPAC Extragalactic Database (NED), 
which is operated by the Jet Propulsion Laboratory, California Institute of Technology, 
under contract with the National Aeronautics and Space Administration.
This work was partly supported by the European Community's Human Potential Programme
under contract HPRN-CT-2002-00321 (ENIGMA). The St.\ Petersburg team  
acknowledges support from Russian Federal Program for Basic Research under grant 05-02-17562.
RATAN-600 observations were partly supported by the
Russian Foundation for Basic Research grant 05-02-17377.
C.\ D\'{\i}az acknowledges support from the Spanish Programa Nacional de
Astronom\'{\i}a y Astrof\'{\i}sica under grant AYA2003-1676.
\end{acknowledgements}

%\bibliographystyle{aa}
%\bibliography{/home/raiteri/my.bib}

\end{document}